\documentclass[preprint,showpacs,preprintnumbers,amsmath,amssymb]{revtex4}

\usepackage{graphicx}% Include figure files
\usepackage{dcolumn}% Align table columns on decimal point
\usepackage{bm}% bold math

\begin{document}

\preprint{PRESAT-8401}

\title{First-Principles Study for Evidence of Low Interface Defect Density at Ge/GeO$_2$ Interfaces }% Force line breaks with \\

\author{Shoichiro Saito}
\author{Takuji Hosoi}
\author{Heiji Watanabe}
\author{Tomoya Ono}

\affiliation{%
Graduate School of Engineering, Osaka University, 2-1 Yamadaoka, Suita, Osaka 565-0871, Japan }%

\date{\today}% It is always \today, today,

\begin{abstract}
We present the evidence of the low defect density at Ge/GeO$_2$ interfaces in terms of first-principles total energy calculations. The energy advantages of the atom emission from the Ge/GeO$_2$ interface to release the stress due to the lattice mismatch are compared with those from the Si/SiO$_2$ interface. The energy advantages of the Ge/GeO$_2$ are found to be smaller than those of the Si/SiO$_2$ because of the high flexibility of the bonding networks in GeO$_2$. Thus, the suppression of the Ge-atom emission during the oxidation process leads to the improved electrical properties of the Ge/GeO$_2$ interfaces.
\end{abstract}

\maketitle

For the past several decades, the performance of Si/SiO$_2$ based transistor has been mainly accomplished by scaling the transistor dimensions resulting in a higher circuit speed, higher packing densities, and less power consumption. At present, the aggressive scaling of the SiO$_2$ gate dielectric in silicon metal-oxide-semiconductor (MOS) devices leads to unacceptably high gate leakage currents. Thus, the scaling of advanced devices is approaching its technological and fundamental limits. To further improve the performance of complementary MOS devices, alternative channel materials are considered. Germanium is introduced as one of the promising candidates for beyond scaling devices, because the intrinsic carrier mobility of germanium is higher than that of silicon. To realize high-performance devices with germanium channel, one of the most crucial issues is the formation of gate stacks with superior interface properties. Ge/GeO$_2$ interface is one of the most important issue because it exists even in Ge/high-\textit{k} oxide interfaces. Although one might expect that germanium, as another elemental column-IV semiconductor, would behave similarly to silicon, the Ge/GeO$_2$ interface had been generally considered to be more defective than the Si/SiO$_2$ interface, in which the interface trap density typically lie in the range of the latter half of 10$^{11}$ $\sim$ 10$^{12}$ cm$^{-2}$eV$^{-1}$ \cite{fukuda}. Very recently, Matsubara \textit{et al}. \cite{takagi} have reported that the minimum value of the interface trap density lower than 10$^{11}$ cm$^{-2}$eV$^{-1}$ can be obtained for Ge/GeO$_2$ MOS interfaces fabricated by dry oxidation without any hydrogen passivation treatment. In addition, the current authors' group (T. H. and H. W.) also accomplished high-quality Ge/GeO$_2$ interfaces by conventional thermal treatment of germanium substrates \cite{hosoi}. 

On the theoretical side, the structural and electronic properties of the Ge/GeO$_2$ interface have been investigated \cite{Pourtois,houssa,houssa2}. Houssa \textit{et al}. \cite{houssa2} simulated the density of germanium dangling bonds at the Ge/GeO$_2$ interface as a function of the oxidation temperature by combining viscoelastic data of GeO$_2$ and the modified Maxwell's model, and claimed that the density of germanium dangling bonds is less than that of silicon dangling bonds. Their results are in the good agreement with the experiments done by Matsubara \textit{et al}. \cite{takagi} and our result \cite{hosoi}. The future development of the passivation technologies to deactivate the interface defects, by introducing terminators such as hydrogen, sulfur, silicon, and so on, will make it possible to achieve the lower interface defect density in Ge/GeO$_2$. However, the formation of the defects in the Ge/GeO$_2$ interface region during the oxidation process is not well-known. 

In this Letter, we implement the first-principles total energy calculation of the Ge/GeO$_2$ interface during the oxidation process. In the case of the Si/SiO$_2$ interface, Kageshima and Shiraishi \cite{kageshima} simulated oxidation processes by first principles. They found that a quartz structure can be obtained if silicon atoms are kicked out from the interface during oxidation, and these emitted silicon atoms might result in the creation of the interface defects. We compared the emission probabilities of the atom from the Ge/GeO$_2$ and Si/SiO$_2$ interfaces in terms of the total energy calculation following the oxidation process proposed by Kageshima and Shiraishi. To the best of our knowledge, this is the first theoretical investigation of the evidence of the low interface defect density at the Ge/GeO$_2$ interfaces by first-principles total energy calculations. We found that the energy advantages of the germanium-atom emission are lower than that of the silicon-atom emission, and the small energy advantages are strongly related to the high flexibility of the bonding networks in GeO$_2$.

%figure
\begin{figure}
\includegraphics{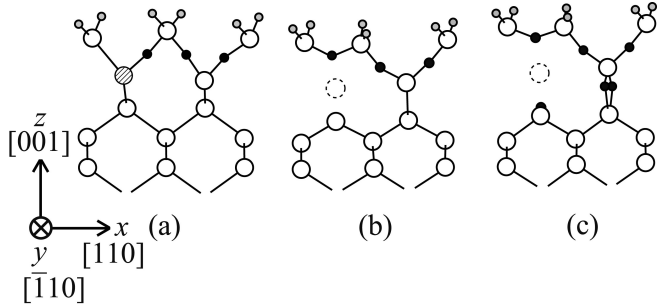}% Here is how to import EPS art
\caption{\label{fig:surfmodel} Atomic configurations of surface models. (a) No germanium atom is emitted from the structure which includes three oxygen atoms. (b) One germanium atom per unit cell is emitted from the structure which includes three oxygen atoms. (c) One germanium atom per unit cell is emitted from the structure which includes six oxygen atoms. The empty circles, the grey circles, and the filled circles are germanium atoms, hydrogen atoms, and oxygen atoms, respectively, and the broken circles indicate the position where the germanium atom is emitted. }
\end{figure}
%figure

%figure
\begin{figure}
\includegraphics{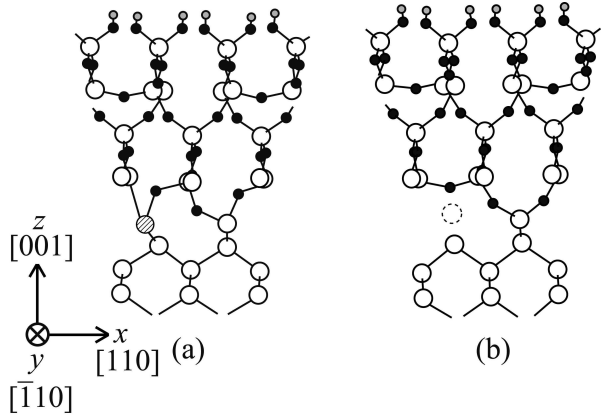}% Here is how to import EPS art
\caption{\label{fig:intfmodel} Atomic configurations of interface models in which additional three oxygen atoms are inserted into the Ge/GeO$_2$ interface region (a) before and (b) after emission. The meanings of the symbols are the same with those in Fig.~ \ref{fig:surfmodel}. }
\end{figure}
%figure

Our calculations were performed based on the density functional theory \cite{ hk,ks} using the real-space finite-difference method \cite{che,ho,oh}. In this method, basis sets used to expand wave functions are not required and boundary conditions are not constrained to be periodic. Our models are imposed the periodic boundary condition in the directions parallel to the surface and the isolated boundary condition in the direction perpendicular to the surface. The norm-conserving pseudopotentials of Troullier and Martins \cite{kb,tm} in the Kleinman-Bylander representation \cite{kobayashi} were employed to describe the electron-ion interaction and the exchange-correlation potential was treated within the local density approximation \cite{pz}. The cutoff energy was set at 112 Ry, which corresponds to a grid spacing of 0.30 a.u., and a further higher cutoff energy was set at 1011 Ry in the vicinity of the nuclei with the augmentation of the double-grid technique \cite{ho,oh}. Eight k-points in the 1 $\times $ 1 lateral unit cell were used for the Brillouin zone sampling. Figure~\ref{fig:surfmodel} shows an example of the Ge(100) surface models. The initial structure of the surface model has nine germanium atomic layers, and the size of the supercell was $L_x = L_y = a_0$ and $L_z = 5.5 a_0$, where $L_x$, $L_y$, and $L_z$ are the lengths of the supercell in the $x$, $y$, and $z$ directions, respectively, and $a_0$ is the experimental lattice constant of the bulk \cite{comment}. Both sides of the surface are simply terminated by hydrogen atoms. We sequentially inserted oxygen atoms between Ge-Ge bonds from the surface, assuming atomical layer-by-layer oxide growth, and finally introduced six oxygen atoms per unit cell. As the further oxidation model, Fig.~\ref{fig:intfmodel} shows an example of the interface models. The initial structure of the interface model has three GeO$_2$ molecule layers and seven germanium atomic layers. The length of the supercell in the $z$ direction was $L_z = 7.1 a_0$, and the other computational details were the same with those described in the surface models. We calculated the interface models inserted three or six oxygen atoms per unit cell into the interface region. During the first-principles structural optimization, we relaxed all the atoms except the germanium atoms in the bottom-most layer and the hydrogen atoms terminating their dangling bonds, reaching a tolerance in the force of $F_{max}<$1.0 mH/bohr. For the comparison of the energy advantages between the cases of germanium and silicon, the same models for Si/SiO$_2$ were also calculated.

%figure
\begin{figure}
\includegraphics{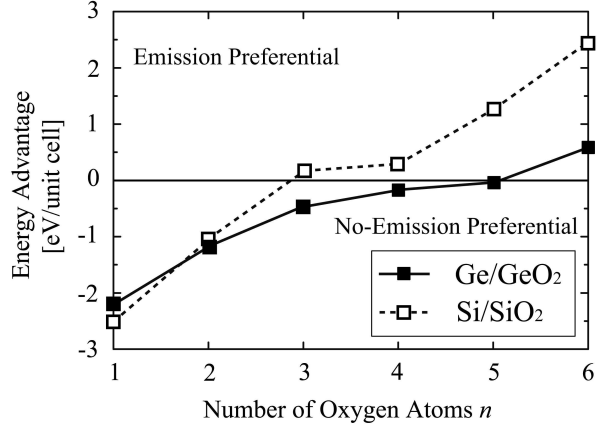}% Here is how to import EPS art
\caption{\label{fig:eagraph} Energy advantage of germanium and silicon emitting structures compared with nonemitting structures as a function of the number of the oxygen atoms per unit cell, \textit{n}. }
\end{figure}
%figure

%table
\begin{table}
\caption{\label{tbl:table1} Energy advantage of germanium and silicon emitting structures compared with nonemitting structures in interface models. $n$ indicates the number of the inserted oxygen atoms at the interface. All the values are in eV/unit cell. }
\begin{ruledtabular}
\begin{tabular}{lcc}
&$n$ = 3 &$n$ = 6 \\
\hline
Quartz/Ge(100) & $-$1.01 & $-$0.31\\
Quartz/Si(100) & $-$0.13 & \ \ 3.94\\
\end{tabular}
\end{ruledtabular}
\end{table}
%table

First, we examine the initial oxidation process of the substrates using the surface models. According to the oxidation process proposed by Kageshima and Shiraishi \cite{kageshima}, the silicon atom in the second layer will be emitted to release the stress. Figures~\ref{fig:surfmodel}(a) and \ref{fig:surfmodel}(b) show the examples for the case of $n = 3$, where $n$ is the number of the inserted oxygen atoms. The hatched silicon atom is kicked out and goes elsewhere. When $n$ is smaller than five, the dangling bonds remain at the interface after the emission. On the contrary, when six oxygen atoms are introduced, these dangling bonds are terminated by forming a Ge-O-Ge bond [see Fig.~\ref{fig:surfmodel}(c)]. Figure~\ref{fig:eagraph} shows the energy advantages of germanium and silicon as a function of the number of the inserted oxygen atoms. The energy advantage is defined as $E_{non}(n)-(E_{emi}(n)+ \mu)$, where $E_{non}(n)$ and $E_{emi}(n)$ are the total energies of the nonemitting and emitted structure with $n$ oxygen atoms inserted, respectively. In addition, $\mu$ is the chemical potential of the germanium or silicon in the bulk phase \cite{mucomment}. The energy advantage of germanium for the atom emission is larger than that of silicon at $n = 1$ because the formation energy of monovacancy is significantly smaller in germanium ($\sim$ 1.9 eV) \cite{vacancy1} than silicon ($\sim$ 3.3 eV) \cite{vacancy2}. On the other hand, the energy advantage of silicon increases more rapidly as the oxygen atoms are inserted and become larger than that of germanium. The silicon-atom emission from the surface model is preferential when $n$ is larger than three, while the germanium-atom emission is only at $n = 6$. We also examine the energy advantage of the atom emission from the interfaces as shown in Fig.~\ref{fig:intfmodel}. Table~\ref{tbl:table1} summarizes the energy advantages for the emission from the interface, which are computed by the same manner with the case of the surface models. The energy advantage of germanium remains smaller than that of silicon. Although the quantitative discussion for the energy advantage is difficult because the chemical potentials of the atoms are simply computed from those in the bulk, our results indicate that the germanium atom preferentially stays in the substrate during the oxidation. In addition, if the germanium atoms are kicked out at $n = 6$, the two remaining dangling bonds are saturated by the reconstruction of the Ge-O-Ge bonding network as mentioned above. Since the event creating the dangling bonds at the interface scarcely happens, the defect density of the Ge/GeO$_2$ interface becomes lower than that of the Si/SiO$_2$ interface.

%table
\begin{table}
\caption{\label{tbl:table2} Dispersion of bonding angles and thickness of oxidized region. The dispersions are calculated as $\sum_{i=1}^{n} (109.5 - \theta_i)^2/n$, where $\theta_i$ indicates the bonding angles around germanium (silicon) atoms, and $n = 24$ corresponding to the bonding angles of atoms in the first and second germanium (silicon) atomic layers per unit cell. The oxide region is the distance between the first and second germanium (silicon) atomic layers. $a_0$ indicates the experimental lattice constants of the bulk phase.}
\begin{ruledtabular}
\begin{tabular}{lcc}
& Dispersion[degree$^2$] & Thickness \\
\hline
Ge & 283.4 & 0.43$a_0$\\
Si & 246.1 & 0.40$a_0$\\
\end{tabular}
\end{ruledtabular}
\end{table}
%table

When oxygen atoms are sequentially inserted between Ge-Ge bonding networks from the surface, the formed oxide region has Ge-O-Ge bonding networks similar to that of the cristobalite of GeO$_2$ bulk, and the bonding networks are attained lattice mismatch of 21 \%, which is almost the same as the lattice mismatch between Si and SiO$_2$. Therefore, the difference in the lattice mismatch is not strongly related to the small energy advantage of the Ge/GeO$_2$ interface. Then, we compare the dispersions of the bonding angles in the GeO$_2$ and SiO$_2$ regions at $n = 4$ which correspond to the 1-atomic-layer-oxidized surface. Table~\ref{tbl:table2} shows the dispersions of the bonding angles from the tetrahedral structure of 109.5 degree and the thickness of the oxidized region. The larger dispersion of the bonding angles around germanium means that the bonding angles around germanium are more flexible than that around silicon. Therefore, the Young modulus of GeO$_2$ is lower than that of SiO$_2$ in the good agreement with the experiment \cite{young}. Since the O-Si-O bonding angles keep the tetrahedral structure rigidly, the thickness of the SiO$_2$ region barely increases to release the stress in the lattice mismatch while the GeO$_2$ region elongates to the vacuum region. Thus, the high flexibility of the O-Ge-O bonding networks causes the small energy advantages of the germanium-atom emission, and then leads to the lower defect density at the Ge/GeO$_2$ interface. Furthermore, the emitted atoms might cause the source of self-interstitials. The creation of these interstitials at the interface or in the substrate also becomes a serious problem because the interstitials cause the degradation of the MOS devices such as the dielectric leakage current and scattering centers of inversion carriers. Since the formation energy of the silicon self-interstitials ($\sim$ 3.5 eV) is almost equal to that of germanium self-interstitials, the large energy advantage of silicon also gives rise to the self-interstitials at the interface \cite{gefe,sife}. Thus, our results imply that the Ge/GeO$_2$ MOS field effect transistor exhibits the greater performance than the Si/SiO$_2$ one owing to the lower interface trap density at its interface as well as high carrier mobility. 

In conclusion, we have investigated the germanium- and silicon-atom emission during the oxidation process using the first-principles calculation. It is revealed that the germanium-atom emission hardly occurs compared with the silicon-atom emission because the high flexibility of the O-Ge-O bonding networks decreases the stress due to the lattice mismatch at the Ge/GeO$_2$ interface. Our study supports the experimental results that the lower interface trap density will be realizable in the Ge/GeO$_2$ interface. Since we studied only limited models of surfaces and interfaces, not all of the details of the actual process are included. Nevertheless, we believe that our study will help to design the germanium based device as the beyond scaling devices. 

The authors would like to thank Professor Kikuji Hirose of Osaka University,  Professor Kenji Shiraishi of University of Tsukuba, and Professor Hiroyuki Kageshima of NTT Basic Research Laboratories for reading the entire text in its original form and fruitful discussion. 
 This research was partially supported by a Grant-in-Aid for the Global COE ``Center of Excellence for Atomically Controlled Fabrication Technology'', by a Grant-in-Aid for Scientific Research in Priority Areas ``Development of New Quantum Simulators and Quantum Design'' (Grant No. 17064012), and also by a Grant-in-Aid for Young Scientists (B) (Grant No. 20710078) from the Ministry of Education, Culture, Sports, Science and Technology. The numerical calculation was carried out using the computer facilities of the Institute for Solid State Physics at the University of Tokyo, the Research Center for Computational Science at the National Institute of Natural Science, and the Information Synergy Center at Tohoku University.


\begin{thebibliography}{l}
\bibitem{comment}
The experimental lattice constants are 10.67 a.u. for germanium and 10.26 a.u. for silicon.
\bibitem{mucomment}
The chemical potentials $\mu$s are $-$3.97 and $-$3.96 a.u./atom for germanium and silicon, respectively.
\end{thebibliography}
\end{document}